\documentclass[10pt]{article}

\usepackage[utf8]{inputenc}

\usepackage{listings} 
\usepackage{epsfig}
\usepackage{graphicx}
\usepackage{amsmath}
\usepackage{amsfonts}
\usepackage{rotating}
\usepackage{authblk}
\usepackage{float}
\usepackage[english]{babel}
\usepackage{setspace}
\usepackage{amssymb}
\usepackage{ragged2e}
\usepackage{xcolor}
\usepackage{footmisc}
\usepackage[a4paper, margin=1in]{geometry}

\newcommand{\be}{\begin{eqnarray}}
\newcommand{\ee}{\end{eqnarray}}
\newcommand{\ba}{\begin{array}}
\newcommand{\ea}{\end{array}}
\newcommand{\bi}{\begin{itemize}}
\newcommand{\ei}{\end{itemize}}
\newcommand{\bv}{\begin{verbatim}}
\newcommand{\ev}{\end{verbatim}}
\newcommand{\bt}{\begin{tabular}}
\newcommand{\et}{\end{tabular}}
\newcommand{\btab}{\begin{table}}
\newcommand{\etab}{\end{table}}
\newcommand{\bfig}{\begin{figure}}
\newcommand{\efig}{\end{figure}}
\newcommand{\bc}{\begin{center}}
\newcommand{\ec}{\end{center}}
\newcommand{\bit}{\begin{itemize}}
\newcommand{\eit}{\end{itemize}}
\newcommand{\nn}{\nonumber}

\newcommand{\tw}{\textwidth}
\newcommand{\ig}[1]{\includegraphics[width=#1\tw]}

\newcommand{\ssp}{\vspace{0.02\tw}}

\newcommand{\arr}{$\rightarrow$ }
\newcommand{\bmnp}{\begin{minipage}}
\newcommand{\emnp}{\end{minipage}}
\newcommand{\twoc}[4]{\bmnp{#1\tw} #2 \emnp \bmnp{#3\tw} #4 \emnp}

\justifying

\title{ExGUtils: A python package for statistical analysis with the ex-gaussian probability density}
\author[a]{C.~Moret-Tatay}
\author[b]{D.~Gamermann}
\author[c]{E.~Navarro-Pardo}
\author[d]{P.~Fernandez de C\'ordoba}

\affil[a]{Departamento de Neuropsicobiolog\'ia, Metodolog\'ia y Psicolog\'ia Social - Facultad de Psicolog\'ia, Magisterio y Ciencias de la Educaci\'on,
Sede de San Juan Bautista. Universidad Cat\'olica de Valencia, San Vicente M\'artir - Av. Guillem de Castro 175, 46008-  Valencia, Spain.}
\affil[b]{Universidade Federal do Rio Grande do Sul (UFRGS) - Instituto de Física, Av. Bento Gonçalves 9500, Porto Alegre, Rio Grande do Sul}
\affil[c]{Department of Developmental and Educational Psychology - Faculty of Psychology, Universitat de Val\`encia. Av. Blasco Ibáñez, 21
46010 - Valencia, Spain.}
\affil[d]{Instituto Universitario de Matem\'atica Pura y Aplicada - IUMPA, Universidad Polit\'ecnica de Valencia, E-46022 Valencia, Spain}

\begin{document}
\maketitle

\abstract{The study of reaction times and their underlying cognitive processes is an important field in Psychology. Reaction times are usually modeled through the ex-Gaussian distribution, because it provides a good fit to multiple empirical data. The complexity of this distribution makes the use of computational tools an essential element in the field. Therefore, there is a strong need for efficient and versatile computational tools for the research in this area. In this manuscript we discuss some mathematical details of the ex-Gaussian distribution and apply the ExGUtils package, a set of functions and numerical tools, programmed for python, developed for numerical analysis of data involving the ex-Gaussian probability density. In order to validate the package, we present an extensive analysis of fits obtained with it, discuss advantages and differences between the least squares and maximum likelihood methods and quantitatively evaluate the goodness of the obtained fits (which is usually an overlooked point in most literature in the area). The analysis done allows one to identify outliers in the empirical datasets and criteriously determine if there is a need for data trimming and at which points it should be done.}

\emph{keywords:} {ex-Gaussian distribution, significance testing, reaction times}

\justifying


\section{Introduction}

The reaction time (RT) has became one of the most popular dependent variables on cognitive psychology. Over the last few decades, much research have been carried out on the problems of focusing exclusively on success or fail trials during the performance of a task, but also to emphasize the importance of RT variables due to their relationship to underlying cognitive processes \cite{mcvay2012drifting, wickelgren1977speed, sternberg1966high,ratcliff2012children}. However, RT has a potential disadvantage: its skewed distribution. One should keep in mind that in order to perform data analysis, it is preferable that the data follows a known distribution. If the distribution is not symmetrical, it is possible to carry out some data transformation techniques (e.g. the Tukey scale for correcting skewness distribution), or to apply some trimming techniques, but with these techniques statistics may be altered (in other words a high concentration of cases in a given range may be favored and as a result, statistics can appear biased). On the one hand, transformations can affect the absolute value of the data or modify the relative distances between data. Moreover, when conducting trimming techniques it is not easy to distinguish noisy data from valid information, or in other words,  to set the limits our boundaries between outliers and extreme data \cite{heathcote1991}. Whether we include or exclude outliers often depends on the reason why they might occur, dealing with the decision to classify them as variability in the measurement or as an experimental error. Another option for the analysis of skewed data is to characterize it with a known skewed distribution. This procedure allows one to determine the probability of an event based on the statistical model used to fit the data. A common problem with this approach is to estimate the parameters that characterize the distribution. In practice, when one wants to find out the probability for an event numerically, a quantified probability distribution is required.

Going back to the point on characterizing the data with a specific distribution, there is one distribution that has been widely employed in the literature when fitting RT data: the exponentially modified Gaussian distribution \cite{balota2004visual, epstein2011evidence, gooch2012reaction, hervey2006reaction, leth2000mean, navarro13, west1999age, west2000age}. This distribution is characterized by three parameters, $\mu$, $\sigma$ and $\tau$. The first and second parameters ($\mu$ and $\sigma$), correspond to the average and standard deviation of the Gaussian component, while the third parameter ($\tau$) is the decay rate of the exponential component. This distribution provides good fits to multiple empirical RT distributions \cite{lacouture2008use, luce86, ratcliff2008diffusion}, however there are currently no published statistical tables available for significance testing with this distribution.

In this article we present a package, developed in Python, for performing statistical and numerical analysis of data involving the ex-Gaussian function. Python is a high-level interpreted language simpler than the traditional S-PLUS \cite{heathcote2004} or PASTIS \cite{cousineau1997} for computations with the ex-Gaussian function. The package presented here is called ExGUtils (from ex-Gaussian Utilities), it comprises functions for different numerical analysis, many of them specific for the ex-Gaussian probability density.

The article is organized as follows: in the next section we present the ex-Gaussian distribution, its parameters and the different ways in which the distribution can be parameterized. Following this, we discuss two fitting procedures usually adopted to fit probability distributions: the least squares and the maximum likelihood. In the third section we present the ExGUtils module and we apply it in order to fit experimental data, evaluate the goodness of the fits and discuss the main differences in the two fitting methods. In the last section we present a brief overview.


\section{The ex-Gaussian distribution and its probability density}

Given a randomly distributed $X$ variable that can assume values between minus infinity and plus infinity with probability density given by the gaussian distribution,

\be
g(x) &=& \frac{1}{\sigma\sqrt{2\pi}}e^{-\frac{1}{2}\left(\frac{x-\mu}{\sigma}\right)^2},
\ee
and a second random $Y$ variable that can assume values between zero and plus infinity with probability density given by an exponential distribution,

\be
h(x) &=& \frac{1}{\tau} e^{-\frac{x}{\tau}},
\ee
let's define the $Z$ variable as the sum of the two previous random variables: $Z = X + Y$.

The gaussian distribution has $\mu$ average and $\sigma$ standard deviation, while the average and standard deviation of the $Y$ variable will be both equal to $\tau$. The $Z$ variable will also be a random variable, whose average will be given by the sum of the averages of $X$ and $Y$ and whose variance will be equal to the sum of the variances of $X$ and $Y$:

\be
M &=& \mu + \tau \label{exg:av}\\ 
S^2 &=& \sigma^2 + \tau^2 \label{exg:std}
\ee

Defined as such, the variable $Z$ has a probability density with the form \cite{exgauss}:

\be
f(x)&=&\frac{1}{2\tau}e^{\frac{1}{2\tau}\left(2\mu+\frac{\sigma^2}{\tau}-2x\right)}\textrm{erfc}\left(\frac{\mu+\frac{\sigma^2}{\tau}-x}{\sqrt{2}\sigma}\right)\label{eq:exg}
\ee
which receives the name of ex-Gaussian distribution (from exponential modified gaussian distribution). The erfc function is the complementary error function. One must be careful, for $\mu$ and $\sigma$ are NOT the distribution average and standard deviation for this distribution, instead the average and variance of the ex-Gaussian distribution is given by Eqs. (\ref{exg:av}-\ref{exg:std}): $M=\mu+\tau$ and $S^2=\sigma^2+\tau^2$. On the other hand, a calculation of the skewness of this distribution results in:

\be
t=\int_{-\infty}^\infty \left(\frac{x-M}{S}\right)^3f(x)\textrm{d}x &=& \frac{2\tau^3}{(\sigma^2+\tau^2)^{\frac{3}{2}}}.
\ee
While the gaussian distribution has null skewness, the skewness of the exponential distribution is exactly equal to two, as a result the skewness of the ex-Gaussian has an upper bound equal to two in the limit $\sigma \ll \tau$ (when the exponential component dominates) and a lower bound equal to zero in the limit $\sigma \gg \tau$ (when the gaussian component dominates).

Let's parameterize the ex-Gaussian distribution in terms of its average $M$, standard deviation $S$ and a new skewness parameter $\lambda=\sqrt[3]{\frac{t}{2}}$. Defined in this way, the $\lambda$ parameter can have values between 0 and 1. Now, defining the standard coordinate $z$ ($z=\frac{x-M}{S}$) one can have the ex-Gaussian distribution normalized for average 0 and standard deviation $1$ in terms of only one parameter, its asymmetry $\lambda$:

\be
f_\lambda(z) &=& \frac{1}{2\lambda}e^{\frac{1}{2\lambda^2}(-2z\lambda-3\lambda^2+1)}\textrm{erfc}\left(\frac{-z+\frac{1}{\lambda}-2\lambda}{\sqrt{2}\sqrt{1-\lambda^2}}\right).
\ee

In figure \ref{fig1}, we show plots for the ex-Gaussian function for different values of the parameter $\lambda$. We should note that for very small values of $\lambda$ (less than around 0.2), the ex-Gaussian is almost identical to the gaussian function (see figure \ref{fig2})\footnote{In this cases, the numerical evaluation of the ex-Gaussian distribution in eq. \ref{eq:exg} becomes unstable and one can without loss (to a precision of around one part in a million) approximate the ex-Gaussian by a gaussian distribution.}.

\bfig
\bc
\ig{1.0}{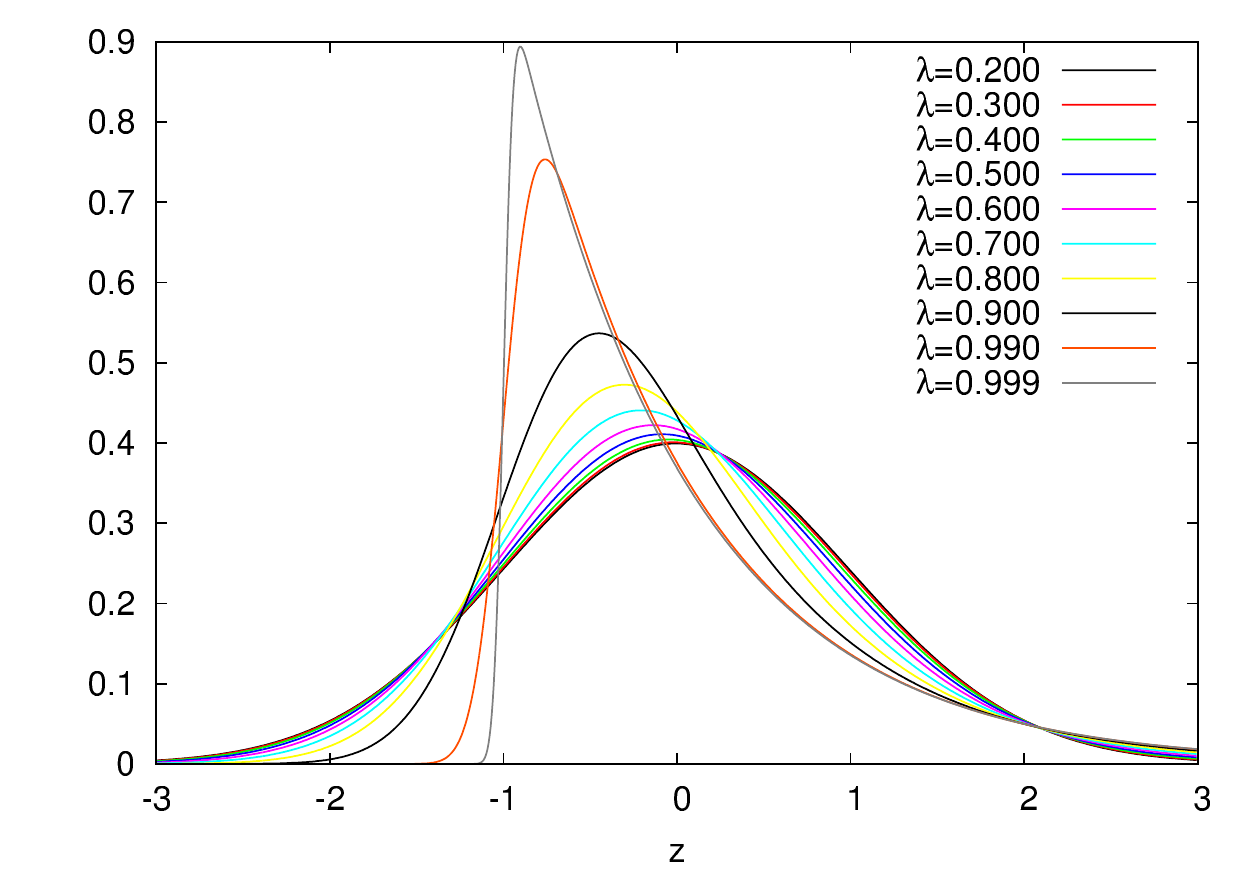}
\ec
\caption{ex-Gaussian distributions for different values of the $\lambda$ asymmetry parameter.}\label{fig1}
\efig

\bfig
\bc
\bt{cc}
\ig{0.4}{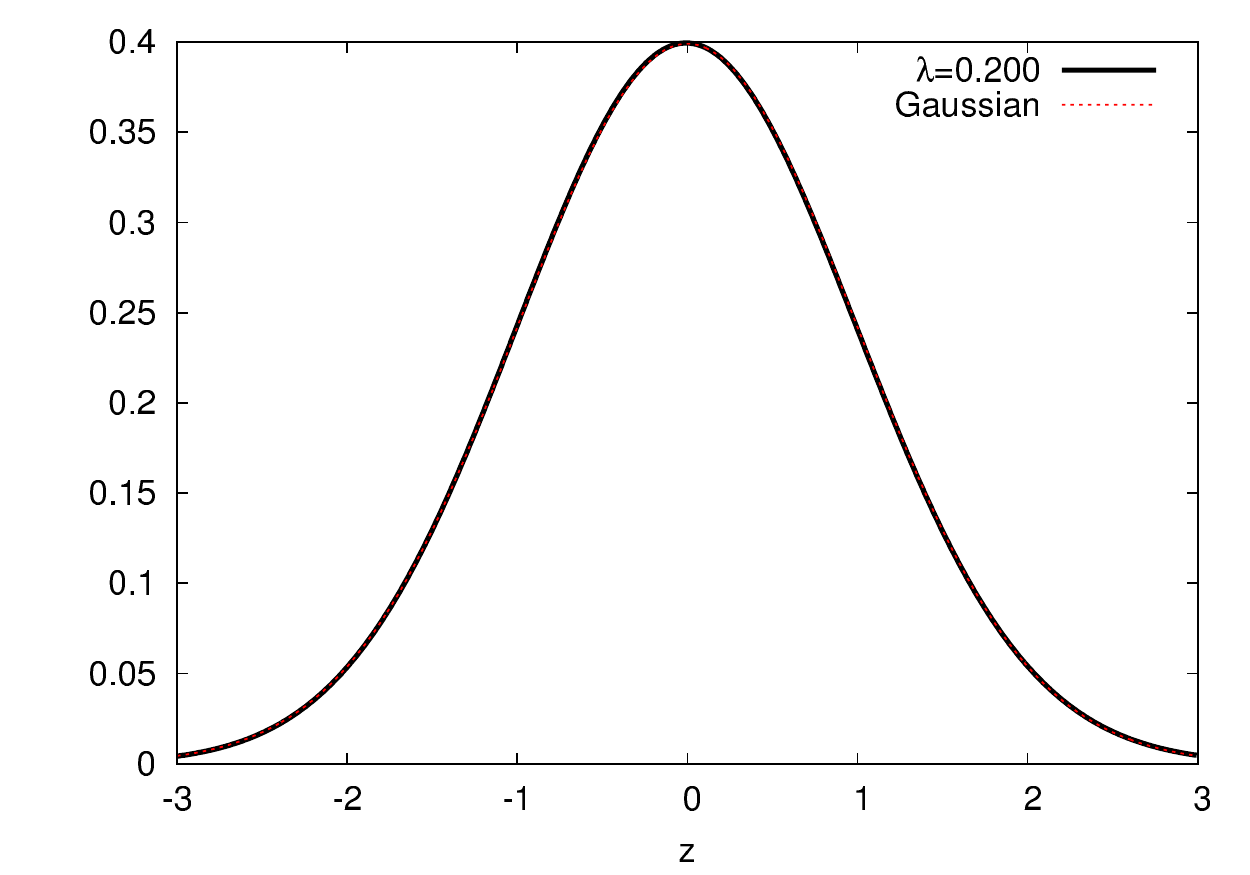} & \ig{0.4}{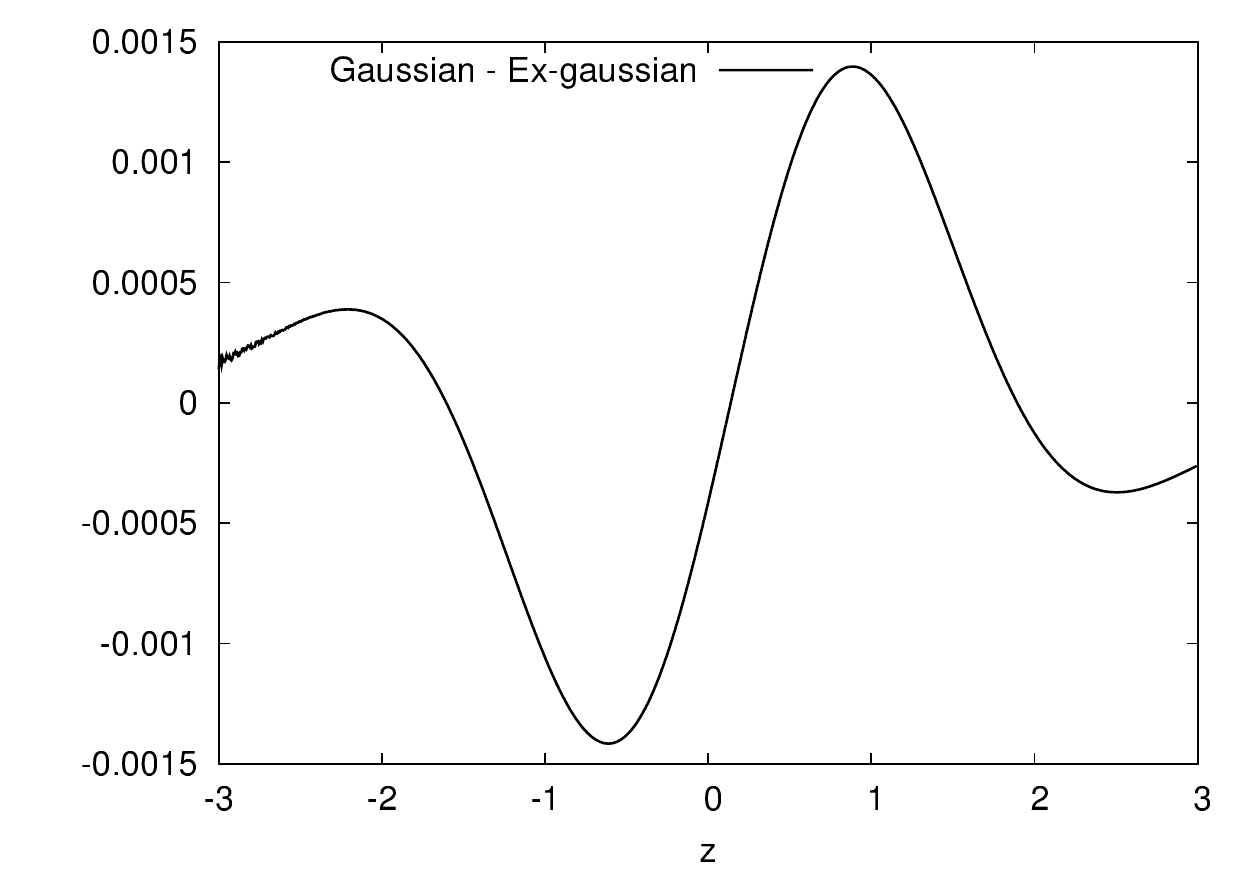}
\et
\ec
\caption{Differences between the ex-Gaussian distribution with $\lambda$=0.2 and the gaussian distribution. Both curves plotted on the left and the difference on the right (note this difference is less than 1\%).}\label{fig2}
\efig

Given a probability density, an important function that can be calculated from it is its cumulative distribution (its left tail), which is the result of the integral

\be
F(z)&=&\int_{-\infty}^z f(x)\textrm{d}x. \label{eq:F}
\ee
The importance of this function is that given the cumulative distribution one is able to calculate the probability of an event. Although the ex-Gaussian distribution has a closed analytical expression, eq. (\ref{eq:exg}), the integral of this expression cannot be analytically evaluated, so one is unable to write an expression for $F(z)$ in eq. (\ref{eq:F}) in the case of the ex-Gaussian and, therefore, numerical methods (computational evaluation) is needed in order to calculate probabilities with the ex-Gaussian distribution.

Let's also define $z_\alpha$, the value of $z$ for which the right tail of the distribution has an area equal to $\alpha$:

\be
\alpha &=& \int_{z_\alpha}^\infty f(x) \textrm{d}x. \label{eq:alpha}
\ee
Eq. (\ref{eq:alpha}) makes even more evident the need of computational resources for the evaluation of $z_\alpha$.


\section{Fitting the Probability Distribution}

We are interested in the following problem: given a dataset, to estimate the parameters $\mu$, $\sigma$ and $\tau$ that, plugged into eq. (\ref{eq:exg}), best fit the data.

We must now define what it means to best fit the data. Different approaches here will result in different values for the parameters. The most trivial approach would be to say that the best parameters are those that result in the theoretical ex-Gaussian distribution with the same statistical parameters: average ($M$), standard deviation ($S$) and asymmetry ($t$ or $\lambda$). So, one can take the dataset, calculate $M$, $S$ and $t$ and use the relations between them and the parameters $\mu$, $\sigma$ and $\tau$:

\hspace{-2.5cm}
\twoc{.5}{
\be
M&=&\mu+\tau\label{conv1}\\
S&=&\sqrt{\sigma^2+\tau^2}\label{conv2}\\
\lambda&=&\sqrt[3]{\frac{t}{2}}=\frac{\tau}{\sqrt{\sigma^2+\tau^2}}\label{conv3}
\ee}{.5}{
\be
\mu&=&M-S\lambda\label{conv4}\\
\sigma&=&S\sqrt{1-\lambda^2}\label{conv5}\\
\tau&=&S\lambda\label{conv6}
\ee}

\ssp

As we will show later, this is the worst possible approach. For instance, in some experiments, one finds the $t$ parameter bigger than 2 (or $\lambda>1$) and from eq. (\ref{conv3}) one sees that, in order to have $t>2$, $\sigma$ cannot be a real number.

Another approach is to find the parameters that minimize the sum of the squared differences between the observed distribution and the theoretical one (least squares). In order to do that, one must, from the dataset, construct its distribution (a histogram), which requires some parametrization (dividing the whole range of observations in fixed intervals). Since a pontentially arbitrary choice is made here, the results might be dependent on this choice. When analyzing data, we will study this dependency and come back to this point.

The last approach we will study is the maximum likelihood method. The function in eq. (\ref{eq:exg}) is a continuous probability distribution for a random variable, which means that $f(x)dx$ can be interpreted as the probability that a observation of the random variable will have the $x$ value (with the infinitesimal uncertainty $dx$). So, given a set of $N$ observations of the random variable, $\{x_i\}$, with $i=1$, $2$, ..., $N$, the likelihood ${\cal L}$ is defined as the probability of such a set, given by:

\be
{\cal L} &=& \prod_{i=1}^N f(x_i; \mu, \sigma, \tau) \label{eq:lkhd}  \\
\ln{\cal L} &=& \sum_{i=1}^N \ln\left(f(x_i; \mu, \sigma, \tau)\right) \label{eq:lnlkhd}
\ee

The maximum likelihood method consists in finding the parameters $\mu$, $\sigma$ and $\tau$ that maximize the likelihood $\cal L$ (or its logarithm\footnote{Note that, since the logarithm is an monotonically increasing function, the maximal argument will result in the maximum value of the function as well.} $\ln\cal L$).  Note that in this approach, one directly uses the observations (data) without the need of any parametrization (histogram). 

In both approaches, least squares and maximum likelihood, one has to find the extreme (maximum or minimum) of a function. The numerical algorithm implemented for this purpose is the steepest descent/ascent (descent for the minimum and ascent for the maximum). The algorithm consists in interactively changing the parameters of the function by amounts given by the gradient of the function in the parameter space until the gradient falls to zero (to a certain precision).


\section{The ExGUtils module}

ExGUtils is a python package. Most of this package is programmed in C, but using the python API that allows one to call the functions from a python interpreter. In this way, one combines the numerical efficiency and precision of the C language with the versatility of the python language. 

The package has two modules: \texttt{pyexg} and \texttt{uts}. The first one comprises all functions with source code programed in python, some of which depend on the \texttt{numpy} and \texttt{scipy} python packages. The module \texttt{uts} on the other hand contains functions with source code programmed in C. In table \ref{tab:funcs} one can find a complete list of all functions in each module and a brief description of it. The source distribution of the ExGUtils module comes with a manual which explains in more detail and with examples each of the functions.

\btab
\caption{Modules and Functions} \label{tab:funcs}
\bc
\bt{c|c|l}
\hline
Module & Function & Brief Description \\
\hline
\hline
   &  drand                 & Returns a random number with homogeneous distribution\\
   &                        &  between 0 and 1. \\
\cline{2-3}
   &  drand\_exp            & Returns a random number with exponential distribution\\
   &                        &  between 0 and infinity.  \\
\cline{2-3}
   &  drand\_gauss          & Returns a random number with gaussian distribution\\
   &                        &  between minus infinity and infinity.  \\
\cline{2-3}
   &  drand\_exg            & Returns a random number with ex-Gaussian distribution\\
   &                        &  between minus infinity and infinity.  \\
\cline{2-3}
   &  gaussian             & Evaluates the gaussian distribution at a given point.\\
\cline{2-3}
   &  exgauss                & Evaluates the ex-Gaussian distribution at a given point.\\
\cline{2-3}
   &  exgauss\_lamb           & Evaluates the ex-Gaussian distribution  parameterized in \\
   &                        &  terms of its asymmetry at a given point. \\
\cline{2-3}
   &  exgauss\_lt                & Evaluates the left-tail of the ex-Gaussian distribution\\
   &                        &   at a given point. \\
\cline{2-3}
   &  exgauss\_lamb\_lt        & Evaluates the left-tail of the ex-Gaussian distribution\\
   &                        &   parameterized in terms of its asymmetry at a given point.  \\
\cline{2-3}
   &  zalp\_exgauss            & Evaluates the point at which the ex-Gaussian distribtion\\
   &                        &  leaves a right-tail equal to $\alpha$.  \\
\cline{2-3}
uts   &  zalp\_exgauss\_lamb  &  Evaluates the point at which the ex-Gaussian distribtion\\
   &                        &    parameterized in terms of its asymmetry leaves a right-tail equal to $\alpha$. \\
\cline{2-3}
   &  pars\_to\_stats           & Given the parameters $\mu$, $\sigma$ and $\tau$, evaluates\\
   &                        &    the corresponding statistics $M$, $S$ and $t$.\\
\cline{2-3}
   &  stats\_to\_pars          & Given the statistics $M$, $S$ and $t$, evaluates the corresponding \\
   &                        &  parameters $\mu$, $\sigma$ and $\tau$.  \\
\cline{2-3}
   &  histogram                & Given a set of observations, produces an histogram.\\
\cline{2-3}
   &  stats                   & Given a set of observations, returns the statistics $M$, $S$ and $t$.\\
\cline{2-3}
   &  stats\_his           & Given a set of observations, presented as a histogram, \\
   &                       & returns the statistics $M$, $S$ and $t$.\\
\cline{2-3}
   &  correlation             & Given a set of points, returns the linear correlation coefficient \\
   &                        &   for the points. \\
\cline{2-3}
   &  minsquare                & Given a set of points, fits a polynomial to the data using \\
   &                        &  the least square method. \\
\cline{2-3}
   &  int\_points\_gauss    & Creates a point partition of an interval for evaluating a\\
   &                        &  gaussian integral.  \\
\cline{2-3}
   &  intsum                & Evaluates the gaussian integral for a function calculated\\
   &                        &  at a gaussian points partition.  \\
\cline{2-3}
   &  exgLKHD                & Evaluates the likelihood and its gradient in the parameter\\
   &                        &   space for a dataset in a given point of the parameter space. \\
\cline{2-3}
   &  maxLKHD                & Evaluates the parameters $\mu$, $\sigma$ and $\tau$ that\\
   &                        &   maximize the likelihood for a given dataset. \\
\cline{2-3}
   &  exgSQR                & Evaluates the sum of squared differences and its gradient in the\\
   &                        &  parameter space for an histogram in a given point of the parameter space.  \\
\cline{2-3}
   &  minSQR                & Evaluates the parameters $\mu$, $\sigma$ and $\tau$ that minimize\\
   &                        &  the sum of squared differences for a given histogram.  \\
\hline
   &  fitter              &  Fits a function using the \texttt{scipy} module. \\
\cline{2-3}
pyexg & zero              &  Finds the zero of an equation.  \\
\cline{2-3}
   &  integral        &  Evaluates an integral.  \\
\cline{2-3}
   &   ANOVA             &  Performs an ANOVA test.    \\
\cline{2-3}
   &   fit\_exgauss       &  Fits an ex-Gaussian to a data set using functions\\
   &                      & in the numpy package.    \\
\hline
\et
\ec
\etab


\section{Applications}

We use here the ExGUtils package in order to analyze data from the experiment in \cite{navarro13}. From this work, we analyse the datasets obtained for the reaction times of different groups of people in recognizing different sets of words in two possible experiments (yes/no and go/nogo).

In our analysis, first each dataset is fitted to the ex-Gaussian distribution through the three different approaches aforementioned:

\bi
\item stat \arr Estimating the parameters through the sample statistics eqs. (\ref{conv4}-\ref{conv6}).
\item minSQR \arr Estimation through least square method, using as initial point in the steepest descent algorithm the $\mu$, $\sigma$ and $\tau$ obtained from the stat method \footnote{\label{fn:assy} In the cases where $t$ was bigger than 2, the inicial parameters were calculated as if $t=1.9$. Note that the final result of the search should not depend on the inicial search point if it starts close to the local maximum/minimum.}.
\item maxLKHD \arr Estimation through maximum likelihood method, using as initial point in the steepest ascent algorithm the $\mu$, $\sigma$ and $\tau$ obtained from the stat method \footref{fn:assy}.
\ei

In table \ref{tab:params}, one can see the estimated parameters and the corresponding statistics for the different experiments. From the table, one sees that in the case of the experiments performed with young people, the value of the skewness, $t$, is bigger than two. This happens because of a few atypical measurements far beyond the bulk of the distribution. In fact, many researches opt for trimming extreme data, by ``arbitrarly'' choosing a cutoff and removing data points beyond this cutoff. One must, though, be careful for the ex-Gaussian distribution does have a long right tail, so we suggest a more criterious procedure:

\begin{sidewaystable}
\small
\caption{Parameters and statistics obtained with the three methods. }\label{tab:params}
\bc
\bt{c||cccccc|cccccc|cccccc}
   & \multicolumn{6}{c|}{stat} & \multicolumn{6}{c|}{minSQR} & \multicolumn{6}{c}{maxLKHD} \\
Experiment & M & S & t & $\mu$ & $\sigma$ & $\tau$  & M & S & t & $\mu$ & $\sigma$ & $\tau$  & M & S & t & $\mu$ & $\sigma$ & $\tau$ \\
\hline
\hline
elder\_gng & 831.14 &  318.95 &  1.75 &  526.06 &  93.02 &  305.08 &  841.45 &  334.97 & 1.85 &  515.16 &  75.76 &  326.29 &  831.14 &  318.39 &  1.79 &  524.49 &  85.67 & 306.65 \\
\hline
elder\_hfgng & 798.55 &  310.00 &  1.94 &  491.52 &  42.78 &  307.03 &  803.42 &  308.32 & 1.82 &  504.66 &  76.18 &  298.76 &  796.63 &  312.32 &  1.90 &  489.39 &  56.11 & 307.24 \\
\hline
elder\_hfyn & 826.15 &  278.61 &  1.62 &  566.56 &  101.16 &  259.59 &  810.03 &  256.41 & 1.68 &  568.03 &  84.74 &  242.00 &  826.15 &  275.10 &  1.75 &  563.06 &  80.40 & 263.08 \\
\hline
elder\_lfgng & 863.73 &  324.53 &  1.60 &  562.65 &  121.14 &  301.08 &  887.69 &  371.88 & 1.90 &  522.22 &  68.76 &  365.46 &  863.73 &  339.88 &  1.87 &  531.43 &  71.38 & 332.30 \\
\hline
elder\_lfyn & 884.53 &  315.93 &  1.59 &  591.97 &  119.27 &  292.55 &  882.16 &  310.04 & 1.77 &  584.33 &  86.12 &  297.84 &  884.53 &  308.13 &  1.62 &  597.54 &  112.18 & 286.98 \\
\hline
elder\_pseudo & 1189.64 &  416.92 &  0.88 &  872.59 &  270.73 &  317.05 &  1233.90 &  518.22 & 1.79 &  734.21 &  137.33 &  499.69 &  1189.64 &  447.89 &  1.60 &  773.74 &  166.24 & 415.90 \\
\hline
elder\_yn & 854.88 &  298.93 &  1.63 &  575.78 &  107.04 &  279.11 &  846.94 &  286.75 & 1.75 &  572.77 &  84.01 &  274.17 &  854.88 &  292.72 &  1.68 &  578.59 &  96.69 & 276.29 \\
\hline
young\_gng & 597.90 &  169.90 &  2.71 &  - &  - &  - &  590.36 &  142.94 & 1.66 &  455.98 &  48.71 &  134.38 &  597.90 &  154.25 &  1.72 &  451.09 &  47.33 & 146.81 \\
\hline
young\_hfgng & 562.94 &  141.88 &  3.04 &  - &  - &  - &  555.47 &  115.19 & 1.54 &  449.84 &  45.93 &  105.63 &  562.94 &  126.30 &  1.65 &  444.47 &  43.77 & 118.47 \\
\hline
young\_hfyn & 621.16 &  176.99 &  3.88 &  - &  - &  - &  610.51 &  128.15 & 1.53 &  493.32 &  51.86 &  117.19 &  621.16 &  148.38 &  1.65 &  482.08 &  51.70 & 139.08 \\
\hline
young\_lf & 632.96 &  187.61 &  2.46 &  - &  - &  - &  625.36 &  161.43 & 1.65 &  473.80 &  55.61 &  151.55 &  632.96 &  173.23 &  1.71 &  468.51 &  54.43 & 164.45 \\
\hline
young\_lfgng & 632.96 &  187.61 &  2.46 &  - &  - &  - &  625.36 &  161.43 & 1.65 &  473.80 &  55.61 &  151.55 &  632.96 &  173.23 &  1.71 &  468.51 &  54.43 & 164.45 \\
\hline
young\_lfyn & 668.57 &  184.88 &  2.10 &  - &  - &  - &  660.23 &  165.16 & 1.60 &  507.06 &  61.77 &  153.18 &  668.58 &  176.56 &  1.70 &  501.24 &  56.31 & 167.34 \\
\hline
young\_pseudo & 722.53 &  190.36 &  2.37 &  - &  - &  - &  718.60 &  175.25 & 1.64 &  554.66 &  61.92 &  163.94 &  722.53 &  180.58 &  1.68 &  552.03 &  59.47 & 170.50 \\
\hline
young\_yn & 644.37 &  182.41 &  2.90 &  - &  - &  - &  635.20 &  148.99 & 1.62 &  496.41 &  54.18 &  138.78 &  644.37 &  164.31 &  1.69 &  488.96 &  53.34 & 155.42 \\
\hline
\et
\ec
\end{sidewaystable}

Having the tools developed in \texttt{ExGUtils}, one can use the parameters obtained in the fitting procedures (either minSQR or maxLKHD) in order to estimate a point beyond which one should find no more than, let's say, 0.1\% of the distribution. In the appendix, the Listing \ref{script1} shows a quick python command line in order to estimate this point in the case of the young\_gng experiment. The result informs us that, in principle, one should not expect to have more than 0.1\% measurements of reaction times bigger than 1472.84 ms if the parameters of the distribution are the ones adjusted by maxLKHD for the young\_gng empirical data. In fact, in this experiment, one has 2396 measurements of reaction times, from those, 8 are bigger than 1472.8 ms (0.33\%). If one now calculates the statistics for the data, removing these 8 outliers, one obtains:

\be
\ba{ccccccc}
\textrm{stats:} & M = 593.80 & S = 154.30 & t = 1.91 & \mu = 441.82 & \sigma = 26.67 & \tau = 151.98 \\
\textrm{minSQR:} & M =  590.11 & S = 142.44 & t = 1.67 & \mu = 455.96 & \sigma = 47.89 & \tau = 134.14 \\
\textrm{maxMLKHD:} & M = 593.80 & S = 148.44 & t = 1.69 & \mu = 453.52 & \sigma = 48.52 & \tau = 140.29
\ea \nn
\ee
In figure \ref{fig:ygngp} one can see the histogram of data plotted along with three ex-Gaussians resulting from the above parameters.

\bfig[h]
\bc
\ig{1.0}{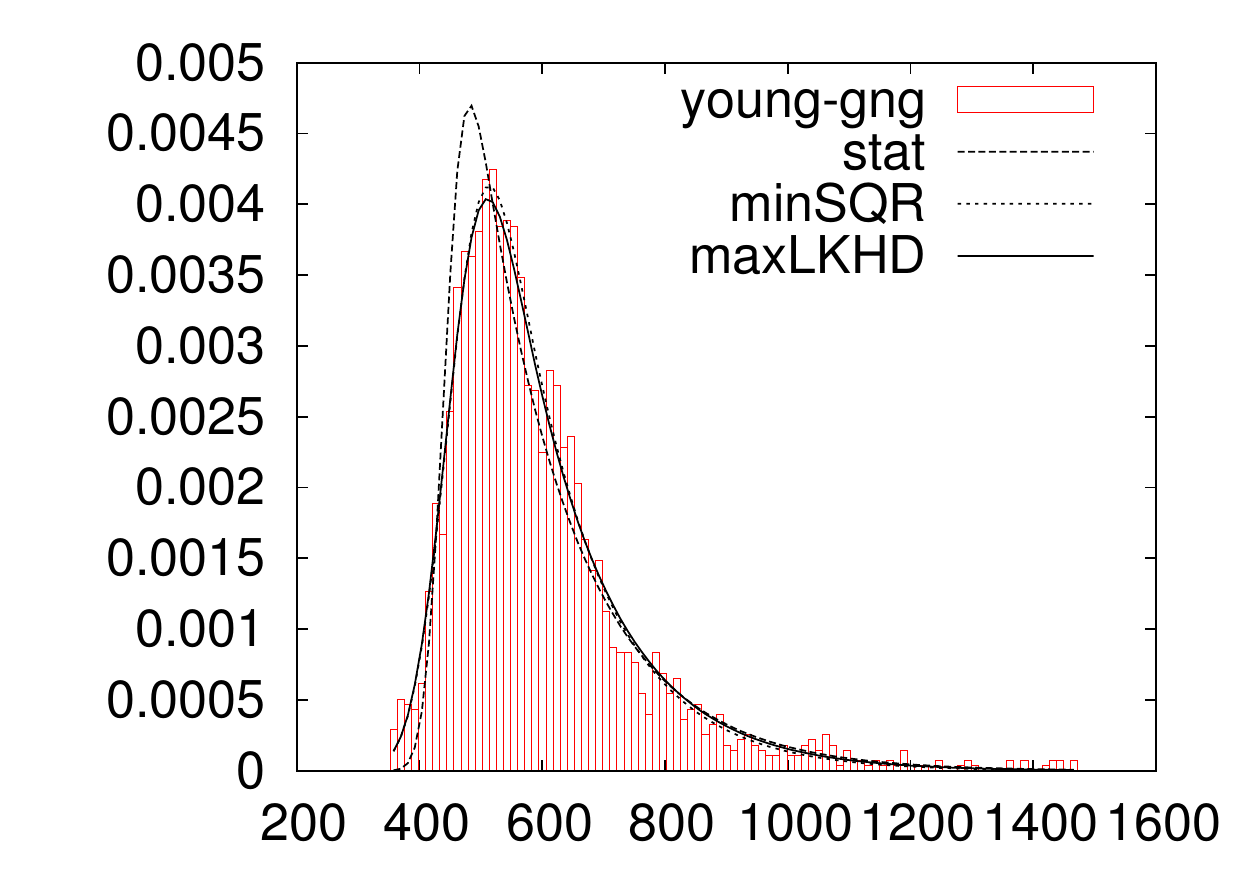}
\ec
\caption{Data for the young\_gng experiment trimmed for outliers with three fitted ex-Gaussians.}\label{fig:ygngp}
\efig

Now, one might ask, having these different fits for the same experiment, how to decide which one is the best? Accepting the parameters of a fit is the same as accepting the null hypothesis that the data measurements come from a population with an ex-Gaussian distribution with the parameters given by the ones obtained by the fit. In \cite{powlaw} the authors suggest a procedure in order to estimate a p-value for this hypothesis when the distribution is a power-law. One can generalize the procedure for any probability distribution, an ex-Gaussian, for example:

\bi
\item Take a measure that quantifies the distance between the data and the fitted theoretical distribution. One could use $\ln \cal L$ or $\chi^2$, but, as our fitting procedures maximize or minimize these measures, as the authors in \cite{powlaw} suggest, in order to avoid any possible bias, we evaluate the Kolmogorov-Smirnov statistic, which can be calculated for reaction-time data without the need of any parametrization.
\item Randomly generate many data samples of the same size as the empirical one using the theoretical distribution with the parameters obtained from the fit to the empirical data.
\item Fit each randomly generated data sample to the theoretical distribution using the same fit procedure used in the case of the empirical data.
\item Evaluate the Kolmogorov-Smirnov statistic between the random sample and its fitted theoretical distribution.
\ei

Following this procedure, one can evaluate the probability that a random data sample, obtained from the fitted distribution, has a bigger distance to the theoretical curve than the distance between the empirical data and its fitted distribution. If this probability is higher than the confidence level one is willing to work with, one can accept the null hypothesis knowing that the probability that one is committing a type I error if one rejects the null hypothesis is $p$.

In the appendix we provide listings with the implementation, in python via the \texttt{ExGUtils} package, of the functions that evaluate this $p$ probability and the Kolmogorov-Smirnov statistic. In table \ref{tab:pvals} we provide the values of $p$ obtained for the experiments, using minSQR and maxLKHD approaches (p1 and p2, respectively).

\btab
\caption{Probabilities p1 and p2 for the fits. KS is the Kolmogorov-Smirnov statistic calculated between the data and its fitted ex-Gaussian. In columns p1 and p2, one finds the probabilities that a randomly generated dataset has a bigger KS statistic than the empirical data. In parenthesis, the average KS statistic and standard deviation for the generated random samples.}\label{tab:pvals}
\bc
\bt{c||cc|cc}
   &  \multicolumn{2}{c|}{minSQR} & \multicolumn{2}{c}{maxLKHD} \\
Experiment & KS & p2 ($\bar{\textrm{KS}}\pm sd$) &  KS  & p1 ($\bar{\textrm{KS}}\pm sd$)   \\
\hline
\hline
elder\_gng & 64.52 & 0.001 (29.47 $\pm$ 8.12) & 38.89 & 0.096 (29.96 $\pm$ 12.54) \\
\hline
elder\_hfgng & 44.32 & 0.001 (20.85 $\pm$ 5.73) & 49.61 & 0.003 (21.33 $\pm$ 5.86) \\
\hline
elder\_hfyn & 34.10 & 0.019 (20.10 $\pm$ 5.35) & 35.30 & 0.021 (20.44 $\pm$ 7.49) \\
\hline
elder\_lfgng & 42.83 & 0.005 (21.73 $\pm$ 5.98) & 31.70 & 0.043 (20.96 $\pm$ 5.94) \\
\hline
elder\_lfyn & 17.25 & 0.634 (19.76 $\pm$ 5.18) & 29.00 & 0.028 (19.15 $\pm$ 5.63) \\
\hline
elder\_pseudo & 62.79 & 0.000 (26.12 $\pm$ 6.81) & 53.10 & 0.009 (25.69 $\pm$ 10.41) \\
\hline
elder\_yn & 32.87 & 0.258 (28.77 $\pm$ 7.42) & 62.72 & 0.012 (29.00 $\pm$ 14.16) \\
\hline
young\_gng & 35.92 & 0.136 (28.60 $\pm$ 7.39) & 69.38 & 0.003 (28.66 $\pm$ 8.36) \\
\hline
young\_hfgng & 21.33 & 0.305 (19.70 $\pm$ 4.99) & 34.11 & 0.016 (20.13 $\pm$ 6.16) \\
\hline
young\_hfyn & 29.75 & 0.049 (19.59 $\pm$ 5.04) & 45.20 & 0.009 (19.83 $\pm$ 7.03) \\
\hline
young\_lf & 22.06 & 0.318 (20.39 $\pm$ 5.81) & 37.78 & 0.015 (20.67 $\pm$ 7.82) \\
\hline
young\_lfgng & 22.06 & 0.299 (20.08 $\pm$ 5.25) & 37.78 & 0.012 (20.27 $\pm$ 6.52) \\
\hline
young\_lfyn & 23.62 & 0.182 (19.66 $\pm$ 5.03) & 17.66 & 0.542 (19.56 $\pm$ 7.43) \\
\hline
young\_pseudo & 20.35 & 0.867 (27.86 $\pm$ 7.20) & 28.48 & 0.386 (28.44 $\pm$ 10.87) \\
\hline
young\_yn & 38.34 & 0.097 (28.07 $\pm$ 7.03) & 54.20 & 0.003 (28.13 $\pm$ 8.66) \\
\hline
\et
\ec
\etab

We can see that there could be some discrepancies in table \ref{tab:pvals}. Sometimes minSQR seems to perform better, sometimes maxLKHD. One might now remember that the minSQR method depends on a parametrization of the data. In order to perform the fit, one needs to construct a histogram of the data, and there is an arbitrary choice in the number of intervals one divides the data into. In the fits performed till now, this number is set to be the default in the \texttt{histogram} function of the \texttt{ExGUtils} package, namely two times the square root of the number of measurements in the data.

In order to study the effect of the number of intervals in the value of p2, we performed the procedure of fitting the data through minSQR after constructing the histogram with different number of intervals. In figure \ref{fig:nints} we show the evolution of the p2 probability, along with the values for $\mu$, $\sigma$ and $\tau$ obtained by minSQR for the histograms constructed with a different number of intervals for the young\_hfgng experiment.

\bfig
\bc
\bt{cc}
\ig{0.4}{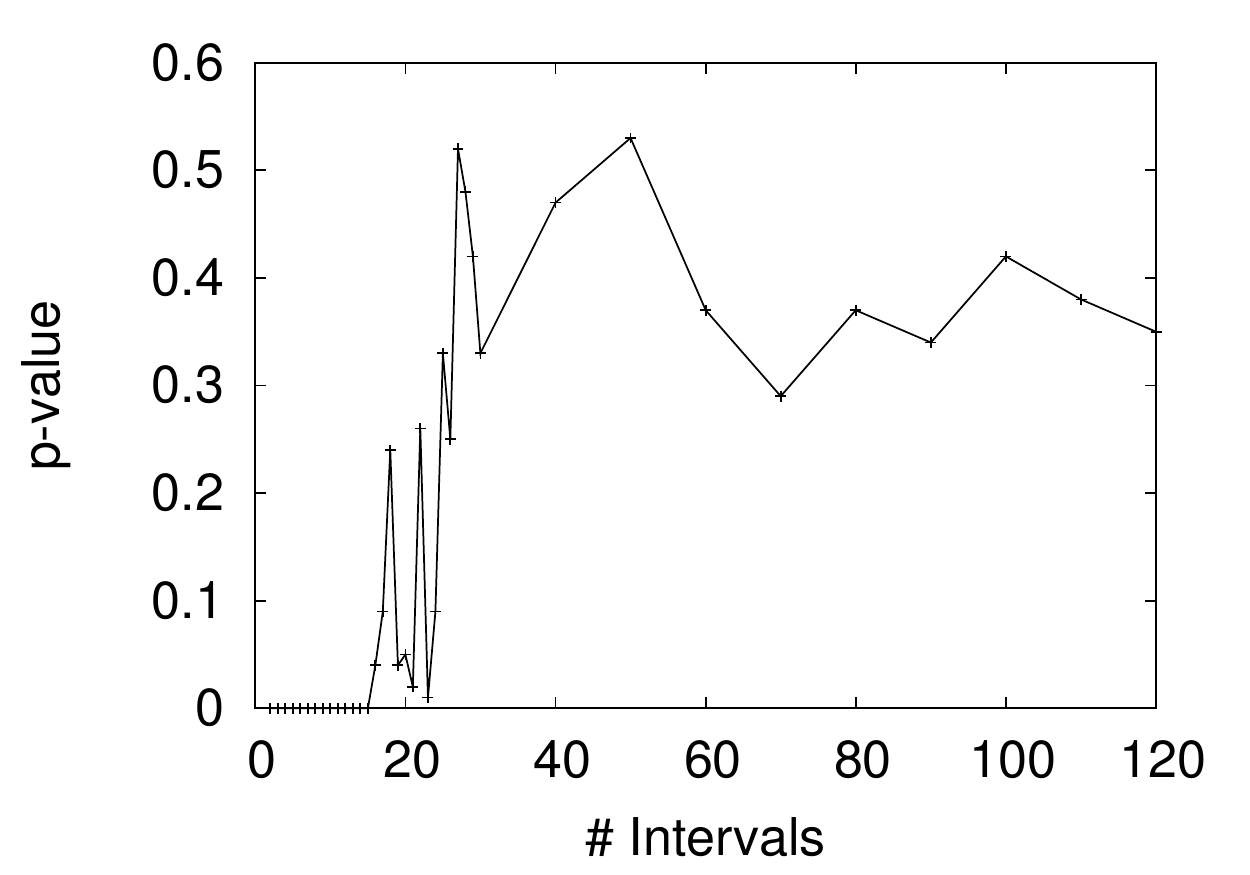} & \ig{0.4}{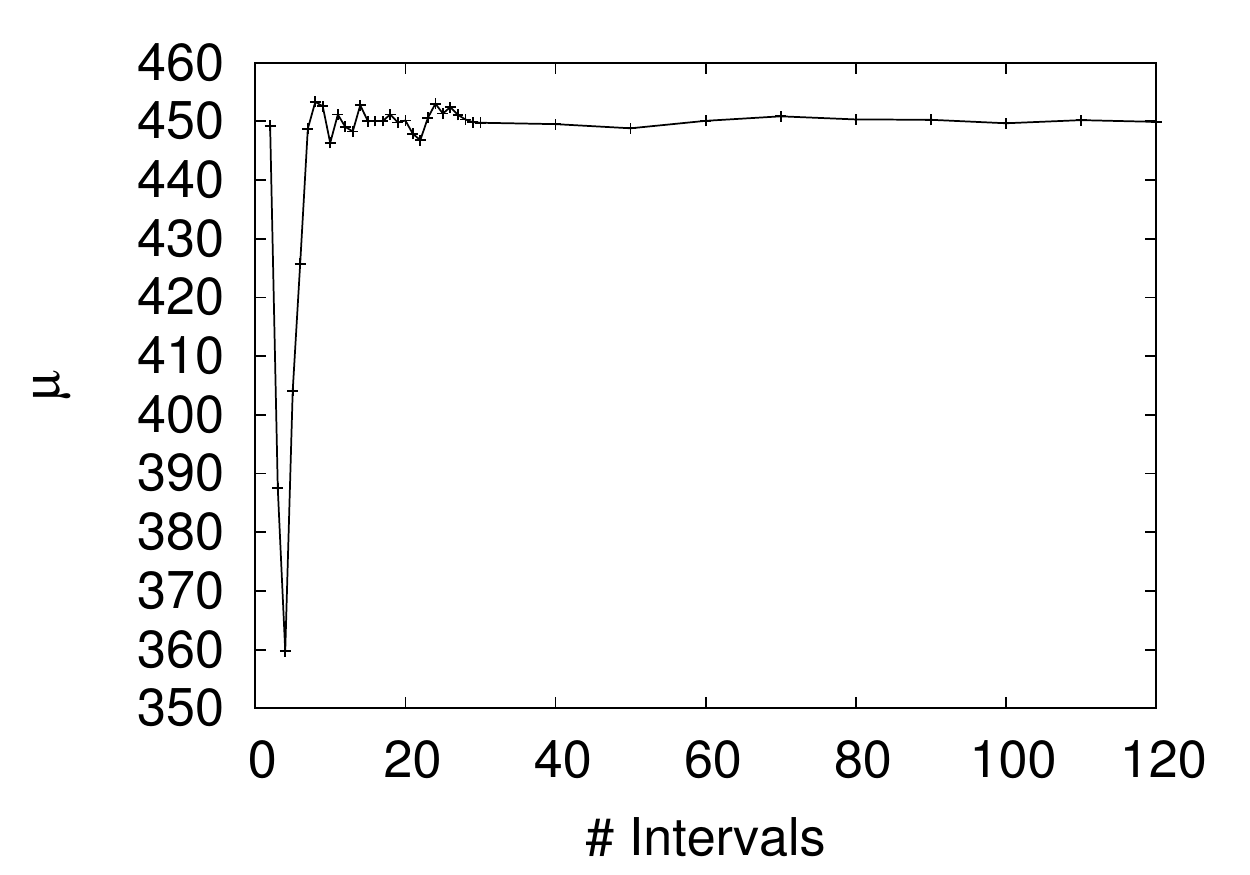} \\
\ig{0.4}{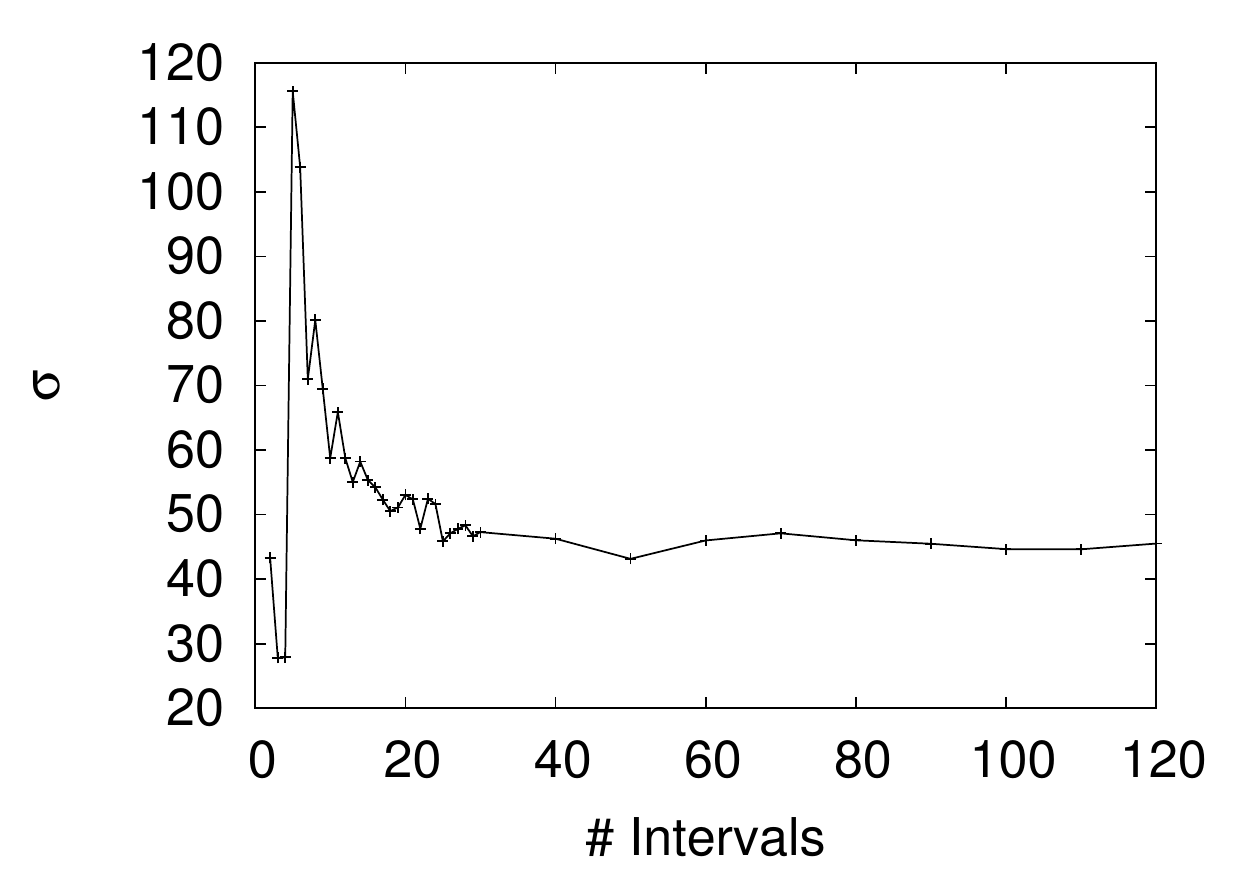} & \ig{0.4}{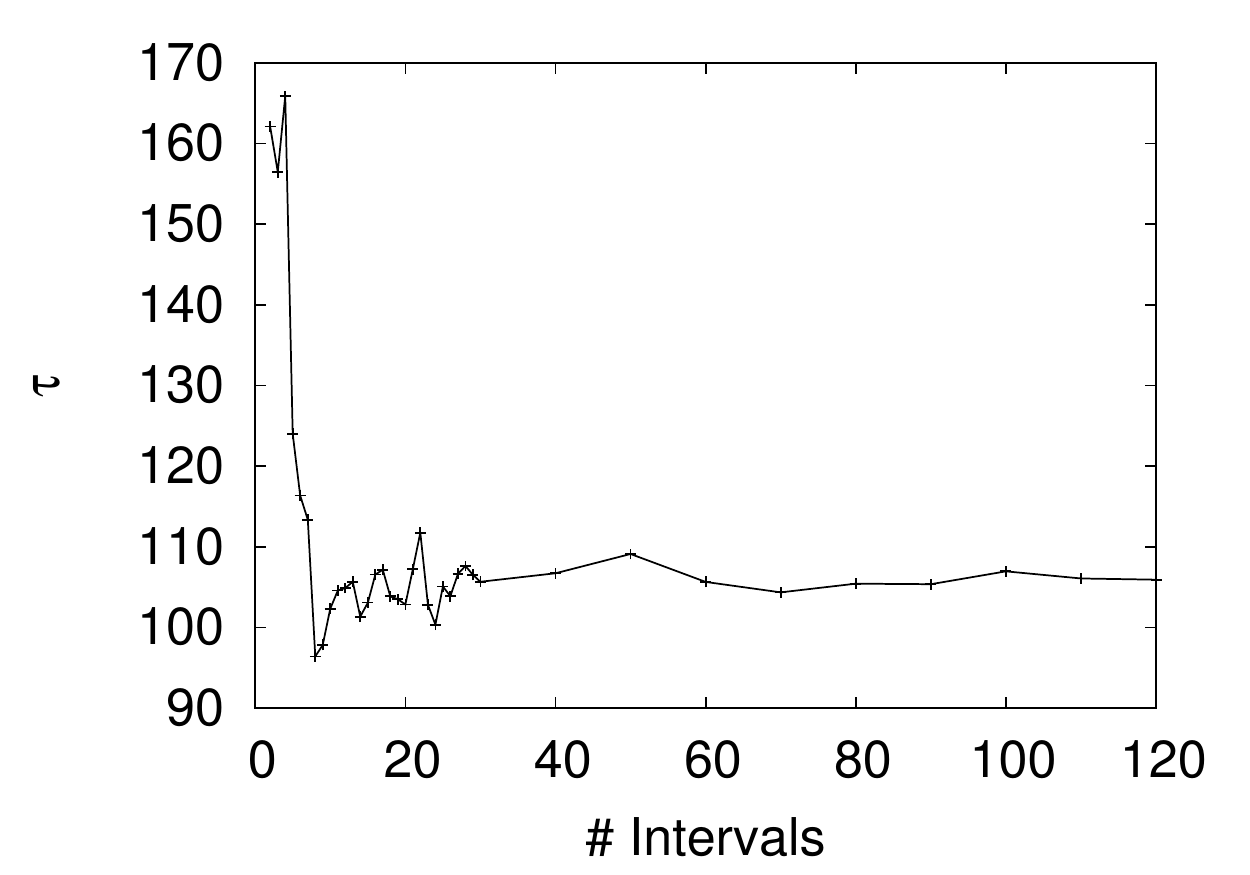} \\
\et
\ec
\caption{Fitting results varying the number of intervals in the histogram for the young\_hfgng experiment. Upper left: evolution of the $p$ probability. Upper right evolution of $\mu$. Bottom left: evolution of $\sigma$. Bottom right: evolution of $\tau$. }\label{fig:nints}
\efig

From the figure one sees that while the number of intervals is unreasonably small compared to the size of the empirical dataset, the values for the fitted ex-Gaussian parameters fluctuate, while the $p$ probability is very small, but, once the number of intervals reaches a reasonable value, around 40, the values for the parameters stabilize and the value of $p$ also gets more stable. So the question remains, why the values for the probability obtained with maxLKHD method is so small in the case of this experiment? The fact is that the likelihood of the dataset is very sensible to outliers. For the value of the probability ($f(x)$ in eq. (\ref{eq:exg})) gets very small for the extreme values. Therefore, in these cases, it might be reasonable to make some criterious data trimming. So we proceed as follows: Given a dataset, we first perform a pre-fitting by maxLKHD. Using the parameters obtained in this fit, we estimate the points where the distribution has a left and right tails of 0.1\% and remove measurements beyond these points. With the trimmed dataset, removed of outliers, we perform fits again and evaluate the p1 and p2 probabilities. In table \ref{tab:pvals2}, we show the results for this new round of fitting and probability evaluations. In more than half of the experiments where one could see a big discrepancy between p1 and p2 in table \ref{tab:pvals}, the trimmed data does show better results. For some datasets, the trimming had no impact on the discrepancy. In any case, one might wonder about the impact of the trimming in the obtained parameters. In table \ref{tab:trim}, we show the results obtained with different trimming criteria.

\btab
\caption{The p1 and p2 probabilities for the fits. KS is the Kolmogorov-Smirnov statistic calculated between the data and its fitted ex-Gaussian. N is the number of data points in each empirical dataset, N$^\prime$ in the number of points removed by the trimming and in brackets next to it its proportion in relation to the total data. In columns p1 and p2, one finds the probabilities that a randomly generated dataset has a bigger KS statistic than the empirical data. In parenthesis, the average KS statistic and standard deviation for the generated random samples.}\label{tab:pvals2}
\bc
\bt{c||cc|cc|cc}
   &  & &  \multicolumn{2}{c|}{minSQR} & \multicolumn{2}{c}{maxLKHD} \\
Experiment & N & N$^\prime$ (\%) & KS & p2 ($\bar{\textrm{KS}}\pm sd$) &   KS  & p1 ($\bar{\textrm{KS}}\pm sd$)  \\
\hline
\hline
elder\_gng & 2348 & 2 (0.09) & 66.58 & 0.000 (28.92 $\pm$ 7.32) & 50.24 & 0.040 (30.98 $\pm$ 17.55) \\
\hline
elder\_hfgng & 1174 & 8 (0.68) & 34.20 & 0.040 (20.67 $\pm$ 5.70) & 32.64 & 0.010 (20.66 $\pm$ 5.83) \\
\hline
elder\_hfyn & 1175 & 2 (0.17) & 32.09 & 0.040 (20.01 $\pm$ 4.86) & 24.76 & 0.090 (19.22 $\pm$ 6.69) \\
\hline
elder\_lfgng & 1174 & 1 (0.09) & 43.49 & 0.000 (21.47 $\pm$ 5.83) & 33.22 & 0.030 (20.57 $\pm$ 6.90) \\
\hline
elder\_lfyn & 1139 & 4 (0.35) & 19.97 & 0.550 (20.55 $\pm$ 6.37) & 19.71 & 0.620 (19.97 $\pm$ 6.11) \\
\hline
elder\_pseudo & 1910 & 5 (0.26) & 57.26 & 0.000 (26.91 $\pm$ 6.64) & 57.11 & 0.010 (26.61 $\pm$ 10.06) \\
\hline
elder\_yn & 2314 & 5 (0.22) & 36.83 & 0.240 (28.57 $\pm$ 7.46) & 29.72 & 0.230 (30.54 $\pm$ 14.33) \\
\hline
young\_gng & 2396 & 10 (0.42) & 38.93 & 0.250 (27.82 $\pm$ 6.32) & 43.11 & 0.020 (30.19 $\pm$ 17.07) \\
\hline
young\_hfgng & 1200 & 8 (0.67) & 23.28 & 0.780 (19.25 $\pm$ 4.39) & 17.82 & 0.430 (18.07 $\pm$ 4.13) \\
\hline
young\_hfyn & 1180 & 9 (0.76) & 27.97 & 0.050 (19.68 $\pm$ 4.91) & 28.93 & 0.010 (20.74 $\pm$ 7.71) \\
\hline
young\_lf & 1196 & 5 (0.42) & 25.11 & 0.310 (20.09 $\pm$ 5.21) & 25.32 & 0.020 (19.69 $\pm$ 4.29) \\
\hline
young\_lfgng & 1196 & 5 (0.42) & 25.11 & 0.280 (20.51 $\pm$ 5.08) & 25.32 & 0.080 (20.55 $\pm$ 5.05) \\
\hline
young\_lfyn & 1132 & 3 (0.27) & 25.20 & 0.230 (19.42 $\pm$ 5.40) & 16.60 & 0.780 (20.72 $\pm$ 8.53) \\
\hline
young\_pseudo & 2326 & 10 (0.43) & 23.33 & 0.940 (27.59 $\pm$ 7.05) & 25.85 & 0.870 (28.45 $\pm$ 12.48) \\
\hline
young\_yn & 2312 & 12 (0.52) & 46.10 & 0.130 (27.80 $\pm$ 7.87) & 28.58 & 0.210 (31.21 $\pm$ 19.74) \\
\hline
\et
\ec
\etab

\btab
\caption{Results for different trimming on the data. The column \% indicates the amount of tail trimmed to the left and right of the data.}\label{tab:trim}
\bc
\bt{cc||cccc|cccc}
   &  &  \multicolumn{4}{c|}{minSQR} & \multicolumn{4}{c}{maxLKHD} \\
Experiment & \% & $\mu$ & $\sigma$ & $\tau$ & p2  & $\mu$ & $\sigma$ & $\tau$ & p1   \\
\hline
\hline
elder\_gng & 0.1 & 513.52  & 73.00  & 329.54 & 0.001  & 518.71  & 75.02  & 313.04 & 0.026  \\
elder\_gng & 0.5 & 516.62  & 76.61  & 319.50 & 0.002  & 521.83  & 70.31  & 299.00 & 0.011  \\
elder\_gng & 1.0 & 516.04  & 76.80  & 317.93 & 0.000  & 523.84  & 66.32  & 291.17 & 0.014  \\
\hline
elder\_hfgng & 0.1 & 509.10  & 84.96  & 285.05 & 0.043  & 504.96  & 65.26  & 297.06 & 0.012  \\
elder\_hfgng & 0.5 & 509.39  & 89.51  & 277.28 & 0.020  & 511.19  & 65.09  & 277.33 & 0.020  \\
elder\_hfgng & 1.0 & 508.40  & 83.49  & 279.35 & 0.016  & 512.79  & 59.89  & 272.67 & 0.005  \\
\hline
elder\_hfyn & 0.1 & 564.82  & 82.19  & 246.63 & 0.052  & 558.93  & 71.17  & 266.45 & 0.148  \\
elder\_hfyn & 0.5 & 565.70  & 83.88  & 242.73 & 0.036  & 559.98  & 68.60  & 261.73 & 0.143  \\
elder\_hfyn & 1.0 & 566.73  & 87.05  & 235.38 & 0.006  & 561.88  & 65.77  & 255.95 & 0.094  \\
\hline
elder\_lfgng & 0.1 & 521.64  & 62.39  & 368.34 & 0.006  & 530.64  & 68.95  & 333.51 & 0.041  \\
elder\_lfgng & 0.5 & 523.29  & 67.46  & 359.50 & 0.006  & 530.25  & 60.81  & 329.35 & 0.011  \\
elder\_lfgng & 1.0 & 523.37  & 67.70  & 356.20 & 0.002  & 533.09  & 59.45  & 318.33 & 0.008  \\
\hline
elder\_lfyn & 0.1 & 583.03  & 84.58  & 301.15 & 0.562  & 581.72  & 76.56  & 305.56 & 0.577  \\
elder\_lfyn & 0.5 & 584.32  & 86.07  & 296.15 & 0.524  & 584.60  & 78.19  & 296.28 & 0.329  \\
elder\_lfyn & 1.0 & 586.72  & 85.93  & 287.48 & 0.470  & 589.73  & 77.85  & 278.47 & 0.027  \\
\hline
elder\_pseudo & 0.1 & 735.04  & 133.55  & 498.90 & 0.001  & 755.81  & 134.79  & 436.48 & 0.012  \\
elder\_pseudo & 0.5 & 733.65  & 135.57  & 499.00 & 0.001  & 754.68  & 132.25  & 438.02 & 0.017  \\
elder\_pseudo & 1.0 & 732.54  & 135.87  & 498.14 & 0.000  & 752.31  & 124.65  & 442.19 & 0.014  \\
\hline
elder\_yn & 0.1 & 572.16  & 81.99  & 275.26 & 0.251  & 567.87  & 73.30  & 288.63 & 0.280  \\
elder\_yn & 0.5 & 573.64  & 84.34  & 270.01 & 0.373  & 570.72  & 72.30  & 278.01 & 0.378  \\
elder\_yn & 1.0 & 573.82  & 84.87  & 266.60 & 0.246  & 573.48  & 72.59  & 268.80 & 0.159  \\
\hline
young\_gng & 0.1 & 456.35  & 48.59  & 133.40 & 0.292  & 453.37  & 47.60  & 140.66 & 0.013  \\
young\_gng & 0.5 & 456.95  & 47.02  & 132.15 & 0.177  & 456.29  & 43.54  & 134.00 & 0.167  \\
young\_gng & 1.0 & 457.70  & 46.28  & 130.55 & 0.096  & 457.63  & 40.37  & 131.00 & 0.013  \\
\hline
young\_hfgng & 0.1 & 449.79  & 45.31  & 105.15 & 0.707  & 448.42  & 44.89  & 109.02 & 0.565  \\
young\_hfgng & 0.5 & 450.77  & 44.72  & 103.91 & 0.500  & 449.62  & 40.74  & 107.45 & 0.704  \\
young\_hfgng & 1.0 & 451.94  & 44.75  & 101.09 & 0.208  & 451.50  & 37.51  & 103.23 & 0.226  \\
\hline
young\_hfyn & 0.1 & 493.66  & 50.92  & 116.16 & 0.032  & 487.17  & 51.93  & 126.49 & 0.009  \\
young\_hfyn & 0.5 & 494.62  & 50.74  & 114.27 & 0.054  & 488.97  & 51.00  & 122.73 & 0.025  \\
young\_hfyn & 1.0 & 495.77  & 50.10  & 111.55 & 0.083  & 493.08  & 49.40  & 114.69 & 0.170  \\
\hline
young\_lf & 0.1 & 473.36  & 54.44  & 151.84 & 0.287  & 471.09  & 54.85  & 157.76 & 0.037  \\
young\_lf & 0.5 & 474.18  & 55.22  & 148.96 & 0.207  & 474.72  & 51.93  & 148.93 & 0.117  \\
young\_lf & 1.0 & 475.03  & 54.10  & 147.35 & 0.067  & 475.22  & 45.69  & 148.46 & 0.019  \\
\hline
young\_lfgng & 0.1 & 473.36  & 54.44  & 151.84 & 0.290  & 471.09  & 54.85  & 157.76 & 0.054  \\
young\_lfgng & 0.5 & 474.18  & 55.22  & 148.96 & 0.201  & 474.72  & 51.93  & 148.93 & 0.119  \\
young\_lfgng & 1.0 & 475.03  & 54.10  & 147.35 & 0.068  & 475.22  & 45.69  & 148.46 & 0.021  \\
\hline
young\_lfyn & 0.1 & 508.16  & 61.53  & 151.83 & 0.228  & 503.17  & 57.27  & 162.27 & 0.776  \\
young\_lfyn & 0.5 & 508.79  & 62.11  & 148.82 & 0.306  & 506.82  & 56.33  & 153.58 & 0.713  \\
young\_lfyn & 1.0 & 508.92  & 59.52  & 148.67 & 0.278  & 508.72  & 51.89  & 151.43 & 0.545  \\
\hline
young\_pseudo & 0.1 & 555.42  & 63.03  & 161.81 & 0.951  & 555.36  & 60.57  & 162.27 & 0.858  \\
young\_pseudo & 0.5 & 556.11  & 63.54  & 159.16 & 0.364  & 556.92  & 57.17  & 158.77 & 0.194  \\
young\_pseudo & 1.0 & 557.18  & 62.50  & 157.25 & 0.096  & 559.57  & 54.06  & 153.59 & 0.021  \\
\hline
young\_yn & 0.1 & 497.56  & 54.59  & 136.65 & 0.141  & 492.23  & 53.69  & 146.70 & 0.144  \\
young\_yn & 0.5 & 498.05  & 54.18  & 135.23 & 0.374  & 495.25  & 52.33  & 139.85 & 0.605  \\
young\_yn & 1.0 & 498.17  & 53.86  & 134.10 & 0.556  & 496.97  & 50.70  & 136.71 & 0.494  \\
\hline
\et
\ec
\etab

Now, having the full picture, one can realize that some values of $p$ are indeed small, indicating that either the ex-Gaussian distribution is not that good a model in order to fit the empirical results, or there is still some systematic error in the analysis of the experiments. Most of the empirical datasets where one sees very low values of $p$ are with elderly people. These have the $\tau$ parameter much bigger than the $\sigma$ which indicates a very asymmetric distribution with a long right tail. Indeed, a careful analysis of the histograms will show that the tail in these empirical distributions seems to be cut short at the extreme of the plots, so that the limit time in the experiment should be bigger than 2500ms in order to get the full distribution. One might argue that the trimming actually was removing data, but most of the removed points in the trimming of elderly data, was from the left tail and not from the right. This issue will result in the wrong evaluation of the KS statistics, since it assumes that one is dealing with the full distribution. This kind of analysis might guide better experiment designs.


\section{Overview}

The ex-Gaussian fit has turned into one of the preferable options when dealing with positive skewed distributions. This technique provides a good fit to multiple empirical data, such as reaction times (a popular variable in Psychology due to its sensibility to underlying cognitive processes). Thus, in this work we present a python package for statistical analysis of data involving this distribution.

This tool allows one to easily work with this alternative strategy (fitting procedure) to the traditional analysis techniques like trimming techniques. This is an advantage given that an ex-Gaussian fit includes all data while trimming may result in biased statistics because of the cuts.

Moreover, this tool is programmed as Python modules, which allow the researcher to integrate them with any other Python resource available. They are also open-source and free software which allows one to develop new tools using these as building blocks.


\section{Availability}

ExGUtils may be downloaded from the Python Package index (https://pypi.python.org/pypi/ExGUtils/2.0) for free along with the source files and the manual with extended explanations on the functions and examples.


\appendix

\section{Python scripts and functions}

Here one finds examples of code programed in python using functions from the \texttt{ExGUtils} package in order to implement some of the methods discussed in the text.

In listing \ref{script1}, one finds a quick command line in order to estimate the cutoff point where one expect to find less than 0.1\% of a sample obtained from an ex-Gaussian distribution.

In listing \ref{ksfunc}, two functions are implemented to evaluate, from a data sample, its KS statistic in relation to a known ex-Gaussian distribution.

In listing \ref{pval1}, a function is implemented in order to, from a list containing numerical data, evaluate the probability that a random sample generated from an ex-Gaussian distribution with know parameters, after being fitted to an ex-Gaussian distribution through the maxLKHD method will have a bigger KS statistic than the empirical data. In listing \ref{pval2} the same function is implemented but the fits are done using the minSQR method.

\lstset{language=python}
\lstset{numbers=left, stepnumber=1}
\begin{lstlisting}[language=Python, firstline=1, lastline=3, label=script1, caption={Determining a cutoff point.}]
>>> from ExGUtils.uts import *
>>> zalp_exgauss(0.999, 451.09, 47.33, 146.81)
1472.8468996267395
\end{lstlisting}

\lstset{language=python}
\lstset{numbers=left, stepnumber=1}
\begin{lstlisting}[language=Python, firstline=1, lastline=13, label=ksfunc, caption={Functions to calculate the Kolmogorov-Smirnov statistic.}]
def CDF(data):
    x = list(set(data))
    x.sort()
    y = [data.count(ele) for ele in x]
    y = [sum(y[0:ii])+1 for ii in xrange(len(y))]
    return x, y

def KS_stat(datas, mu, sig, tau, tot):
    x, y = CDF(datas)
    y2 = [tot*exgauss_lt(ele, mu, sig, tau) for ele in x]
    N = len(y2)
    diffs = [abs(y[ii]-y2[ii]) for ii in xrange(N)]
    return max(diffs)
\end{lstlisting}

\lstset{language=python}
\lstset{numbers=left, stepnumber=1}
\begin{lstlisting}[language=Python, firstline=15, lastline=27, label=pval1, caption={Function to find the probability p1 (fitting done with maxLKHD).}]
def p1(xi, mu, sig, tau, reps=1000, eps=1.e-10):
    N = len(xi)
    xxx = 1./reps
    KSemp = KS_stat(xi, mu, sig, tau, N)
    pval = 0.
    kss = [0. for ii in xrange(reps)]
    for ii in xrange(reps):
        datas = [drand_exg(mu, sig, tau) for jj in xrange(N)]
        nmu, nsig, ntau = maxLKHD(datas, eps=eps)
        KSrand = KS_stat(datas, nmu, nsig, ntau, N)
        kss[ii] = KSrand
        if KSrand>KSemp: pval += xxx
    return pval, stats(kss)
\end{lstlisting}

\lstset{language=python}
\lstset{numbers=left, stepnumber=1}
\begin{lstlisting}[language=Python, firstline=30, lastline=43, label=pval2, caption={Function to find the probability p2 (fitting done with minSQR).}]
def p2(xi, mu, sig, tau, reps=1000, eps=1.e-10):
    N = len(xi)
    xxx = 1./reps
    KSemp = KS_stat(xi, mu, sig, tau, N)
    pval = 0.
    kss = [0. for ii in xrange(reps)]
    for ii in xrange(reps):
        datas = [drand_exg(mu, sig, tau) for jj in xrange(N)]
        [x, y] = histogram(datas, norm=1); 
        nmu, nsig, ntau = minSQR(x, y, mu, sig, tau, eps=eps)
        KSrand = KS_stat(datas, nmu, nsig, ntau, N)
        kss[ii] = KSrand
        if KSrand>KSemp: pval += xxx
    return pval, stats(kss)
\end{lstlisting}


\end{document}